\documentclass[aps,prd,superscriptaddress,groupedaddress,nofootinbib,preprint,longbibliography]{revtex4-1}

\usepackage{asymptote} 
\usepackage{amsmath,amssymb,amsfonts}
\usepackage{color}
\usepackage{xcolor}
\definecolor{red}{rgb}{1,0,0}
\usepackage[colorlinks=true, linkcolor=blue, urlcolor=magenta, citecolor=red]{hyperref}
\usepackage{graphicx}
\graphicspath{{./pic/}}

\newcommand{\bw}{\begin{widetext}}
\newcommand{\ew}{\end{widetext}}
\newcommand{\be}{\begin{equation}}
\newcommand{\en}{\end{equation}}
\newcommand{\bee}{\begin{equation}}
\newcommand{\ene}{\end{equation}}
\newcommand{\bea}{\begin{eqnarray}}
\newcommand{\ena}{\end{eqnarray}}
\newcommand{\bes}{\begin{subequations}}
\newcommand{\ens}{\end{subequations}}
\newcommand{\bef}{\begin{figure}}
\newcommand{\enf}{\end{figure}}

\def\thefootnote{\fnsymbol{footnote}}

\def\ie{{\it i.e.}}
\def\etc{{\it etc.}}

\def\cali{\mathcal{I}}
\def\calj{\mathcal{J}}

\def\calo{\mathcal{O}}

\def\calt{\mathcal{T}}
\def\calu{\mathcal{U}}

\begin{document}


\title{Noncommutative effects on the fluid dynamics and modifications of the Freidmann equation}

\author{Kai Ma}
\email[Electronic address: ]{makainca@yeah.net}
\affiliation{School of Physics Science, Shaanxi University of Technology, Hanzhong 723000, Shaanxi, China}

\date{\today}

\begin{abstract}
We propose a new approach in Lagrangian formalism for studying the fluid dynamics on noncommutative space. Starting with the Poisson bracket for single particle, a map from canonical Lagrangian variables to Eulerian variables is constructed for taking into account of the noncommutative effects. The advantage of this approach is that the kinematic and potential energies in the Lagrangian formalism continuously change in the infinite limit to the ones in Eulerian formalism, and hence make sure that both the kinematical and potential energies are taken into account correctly. Furthermore, in our approach, the equations of motion of the mass density and current density are naturally expressed into conservative form. Based on this approach, the noncommutative Poisson bracket is introduced, and the noncommutative algebra among Eulerian variables, as well as the noncommutative corrections on the equations of motion are obtained. We find that the noncommutative corrections generally depend on the derivatives of potential under consideration. Furthermore, we find that the noncommutative algebra does modify the usual Friedmann equation, and the noncommutative corrections measure the symmetry properties of the density function $\rho(\vec{z})$ under rotation around the direction $\vec{\theta}$. This characterization results in vanishing corrections for spherically symmetric mass density distribution and potential.
\end{abstract}

\maketitle

\tableofcontents

%
\setcounter{page}{1}
\renewcommand{\thefootnote}{\arabic{footnote}}
\setcounter{footnote}{0}

\section{Introduction}\label{sec:intro}
It is a general recognition in modern physics that non-trivial geometry of the background spacetime can affect the dynamics of particles living in it. Various non-trivial background spacetime has been proposed, and among them the noncommutative geometry plays a spacial role. Noncommutative spacetime modifies the fundamental commutator algebra in the ordinary Quantum Mechanics (QM) as 
\bee\label{eq:ncdefine}
[X_{i}, X_{j}] = i \theta_{ij}\,,
\ene
with a totally anti-symmetric constant tensor $\theta_{ij}$ having dimension of length-squared, and hence evolution of the quantum system receives corrections. Noncommutative property not only appears in new physics models, for instance, string theory embedded in a background magnetic field~\cite{Seiberg:1999vs} and  quantum gravity~\cite{Freidel:2005me, Moffat:2000gr, Moffat:2000fv, Faizal:2013ioa}, but also in real quantum system~\cite{Gamboa:2000yq, Ho:2001aa}. Extensive studies have been devoted to investigate its physical consequences, for instance, breaking of rotational symmetry~\cite{Douglas:2001ba, Szabo:2001kg,Ma:2017fnt}, distortion of energy levels of the atoms~\cite{Chaichian:2000si,Zhang:2004yu,Zhang:2011kr,Zhang:2011zua,Ma:2017rwg,Ramirez:2017pmp}, contributions to the topological phase effects~\cite{Chaichian:2000hy,Chaichian:2001pw,Ma:2016rhk,Ma:2016vac,Ma:2014tua,Rodriguez:2017iti,Fateme:2015eaa}, corrections on the spin-orbital interactions~\cite{Ma:2011gc,Basu:2012td,Deriglazov:2016mhk,Wang:2013kp,Wang:2015cua} as well as speed of relativistic charged particles~\cite{Wang:2017azq,Wang:2017arq,Deriglazov:2015zta,Deriglazov:2015wde}.

Most interestingly, it was pointed out that there is a connection between quantum Hall fluids and noncommutative field theory~\cite{Bahcall:1991an,Susskind:2001fb,ElRhalami:2001xf,Barbon:2001dw,Jackiw:2002tw,Barbon:2007ap,Ma:2016rvw}. This relation were further investigated in a deeper perspective in various systems, for instances, the noncommutative property can appear naturally in the excitations of the lowest Landau level, and becomes dominant when the magnetic field is strong enough~\cite{Jackiw:2001dj,DIAOXin-Feng:2015fra,Gangopadhyay:2014afa,Jackiw:2002pn}. In consideration of the correspondence between Poisson bracket and quantum commutator, noncommutative extended dynamics is also interesting even at the classical level~\cite{Pramanik:2013zy,Pramanik:2012fj,Ghosh:2006cb,Deriglazov:2015bqa}, particularly in the fluid dynamics~\cite{Das:2016hmc,Holender:2011px,Abdalla:2012tt,Alavi:2006sr}. 

Two approaches have been proposed to study the noncommutative effects on the fluid dynamics. In the first one, the Eulerian description of the fluid dynamics is employed, and noncommutative corrections are taken into account by directly applying the Groenwald-Moyal $\star$-product~\cite{Seiberg:1999vs,Douglas:2001ba,Gangopadhyay:2014afa,Banerjee:2009gr,Holender:2011px,Abdalla:2012tt}, which is defined as
\begin{equation}\label{eq:starproduct}
f( \vec{X} ) \star g( \vec{X} )
= \exp{\left[\frac{1}{2} \theta_{ij} \partial_{X_{i}}\partial_{Y_{j}}\right]} f( \vec{X} )g(\vec{Y})\bigg|_{\vec{X}=\vec{Y}},
\end{equation}
for two arbitrary infinitely differentiable functions on the ordinary commutative $R^{3}$ space $f( \vec{X} )$ and $g( \vec{X} )$. The alternative approach works in the Lagrangian description. The bracket algebra between the canonical Lagrangian variables are extended in noncommutative space, and then a map~\cite{Jackiw:2004nm} from the Lagrangian variables to Eulerian variables is applied to obtain the noncommutative corrections in the fluid field theory. 

Recently, a full noncommutative extended algebra among the Euler degrees of freedom, \ie, the density and velocity field variables, was proposed and some implications in cosmology physics was also briefly studied~\cite{Das:2016hmc,Banerjee:2014hca}. However, because the noncommutative effects of the potential was not properly taken into account in Ref.~\cite{Das:2016hmc}, a conclusion of that to first order in noncommutative parameter, the Friedmann equation remains unmodified. In this paper, we propose an alternative approach for the map from the canonical Lagrangian variables to the Eulerian variables. Starting from the single particle picture, the noncommutative effects of the potential are carefully investigated, and the Euler equations are written in the conservative form. We find that the Friedmann equation does receive noncommutative corrections at the leading order of noncommutative parameter $\theta$, and the noncommutative algebra modifies the cosmology physics in an anisotropy and inhomogeneity way.

The paper is organized as follows: in Sec.~\ref{sec:map}, based on the single particle picture, we introduce our map from Lagrangian variables to Eulerian variables, and briefly discuss its relation to the approach in Ref. ~\cite{Das:2016hmc}; in Sec.~\ref{sec:ncFluid}, we show how the noncommutative algebra can be taken into account via the approach given in Sec.~\ref{sec:map}, particularly we focus on the noncommutative effects of the external potential; in Sec.~\ref{sec:ncCosm}, the cosmological implications of the noncommutative effects in  our approach are discussed along the line in Ref.~\cite{Das:2016hmc}; summary and our conclusions are given in the final section, Sec.~\ref{sec:sum}.

%
%

\section{Hamiltonian Framework of Fluid Dynamics}\label{sec:map}
There are two sensible descriptions on the fluid dynamics, Eulerian and Lagrangian descriptions, both are aiming to formulating the equations for mass density, momentum and energy in a fluid. While Eulerian description investigates the physics encoded in an observable $\calo(t, \vec{r})$  at a fixed spatial position $\vec{r}$, position is not an independent variable but instead $\vec{X} = \vec{X} ( t, \vec{r})$ in the Lagrangian description. Both these two approaches have been used to investigate the noncommutative corrections on the ordinary fluid dynamics. However, the later one is more reliable since we can directly use the noncommutative algebra in Eq.~\eqref{eq:ncdefine}. Nevertheless, the later one does not imply any simplification, because we need a map to transform the Lagrangian variables, position and velocity, into Eulerian variables, mass density and current, \etc. In Ref.~\cite{Das:2016hmc}, noncommutative fluid dynamics was investigated by using the map introduced in Ref.~\cite{Jackiw:2004nm}. However, there were some ambiguirities in that approach: 1) the definition of the velocity field involves inverse of an integral of canonical variables, and hence the canonical brackets involving velocity field is hard to define; 2) the Hamiltonian is not defined clearly, and hence the equations of motions looks different in form from the ordinary approach. This also results in that the noncommutative effects of the potential was not considered properly in Ref.~\cite{Das:2016hmc}, which we have mentioned in the Introduction. In this section, we introduce a new approach for the map, and show that how the potential can be taken into account correctly.

In order to employing the Poisson bracket to derive the fluid dynamics, and furthermore to incorporate nontrivial bracket algebra, starting from the single particle picture is necessary to find the correct and convenient map from Lagrangian to Eulerian description. Our formalism for this map is based on the dynamics of single particle, and always starts from a system with finite number of particles, and then derive the relevant expressions and canonical algebras in the limit of infinite number of particles. To start, let us first define the canonical Lagrangian variables, the canonical position variable $\vec{X}$, and the canonical velocity variable $\vec{V}$ (here we do not use $\dot{\vec{X}}$ for the velocity since it is one of the equations of motion), for single particle. The basic Poisson brackets among these variables are defined as follows (here and after we will suppress the explicit time dependence)
\bea
\Big\{ V_{i},\, X_{j} \Big\} &=& \frac{1}{m}\delta_{ij} \,,
\\[3mm]
\Big\{ X_{i},\, X_{j} \Big\} &=& 0\,,
\\[3mm]
\Big\{ V_{i},\, V_{j} \Big\} &=& 0\,,
\ena
where $m$ is the mass of the particle and is used here for normalization. The equations of motion of these variables can be obtained by using above algebra with following Hamiltonian,
\bee
H = \frac{1}{2} m\, \vec{V} \cdot \vec{V} + U(\vec{X})\,,
\ene
where $U(\vec{X})$ is the external potential acting on the particle.
According to the Hamiltonian mechanics one has
\bea
\dot{X}_{i} &=&  \Big\{ H,\, X_{i} \Big\} = V_{i} \,,
\\[3mm]
\dot{V}_{i} &=&  \Big\{ H,\, V_{i} \Big\} = \frac{1}{m}\partial_{X_i} U(\vec{X})\,.
\ena
The above equations are just the classical results from Newton's law for single particle. It is straightforward to extended to the case having finite number of particles (without considering the interactions among these particles). In the limit of infinite number of particles, the algebra become continuous and described as follows,
\bea
\label{eq:VX}
\Big\{ V_{i}( \vec{x}),\, X_{j}( \vec{y}  ) \Big\} &=& \frac{1}{\rho_{0}} \delta_{ij} \delta^{3}( \vec{x} - \vec{y} )\,,
\\[3mm]
\label{eq:XX}
\Big\{ X_{i}( \vec{x}),\, X_{j}( \vec{y}  ) \Big\} &=& 0\,,
\\[3mm]
\label{eq:VV}
\Big\{ V_{i}( \vec{x}),\, V_{j}( \vec{y}  ) \Big\} &=& 0\,,
\ena
where $\rho_{0} =  M/\Omega$ for a system with total mass $M$ and volume $\Omega$, and a set of new variables $\vec{x}$, which are fixed for grading the space of fluid, are introduced to account for the infinity of the system. The corresponding Hamiltonian is
\bee\label{eq:inftyHamiltonian}
H = \int d^{3}\vec{x}\,\rho_{0}\,\bigg( \frac{1}{2}  \vec{V}( \vec{x}) \cdot \vec{V}( \vec{x}) + \frac{1}{m} U(\vec{X}( \vec{x})) \bigg)\,.
\ene
The equations of motion can be obtained in the similar way,
\bea
\dot{X}_{i}( \vec{y}  ) &=&  \Big\{ H,\, X_{i}( \vec{y}  ) \Big\} = V_{i}(\vec{y}) \,,
\\[3mm]
\dot{V}_{i}( \vec{y}  ) &=&  \Big\{ H,\, V_{i}( \vec{y}  ) \Big\} = - \frac{1}{m}\partial_{X_i(\vec{y})} U\Big(\vec{X}(\vec{y})\Big)\,.
\ena
Up to here, one can already investigate, in terms of the canonical variables $\vec{X}( \vec{x})$ and $\vec{V}( \vec{x})$, the physical consequences of any extended algebra. However, in terms of the Eulerian variables, a map is necessary.

In Ref.~\cite{Das:2016hmc}, a map introduced in Ref.~\cite{Jackiw:2004nm} was used, however, the definition of the velocity field involves inverse of an integral of canonical variables, and hence the canonical brackets involving velocity field is hard to define. In this paper, we will directly use the current field, instead of the velocity field, to define the map. Specifically, the mass and current densities are defined in terms of $\vec{X}( \vec{x})$ and $\vec{V}( \vec{x})$, and given as follows,
\bea
\label{eq:fluidDensity:Def}
\rho(\vec{r}) &=& \rho_{0} \int d^{3} \vec{x}\; \delta^{3} \big( \vec{X}(\vec{x}) - \vec{r}\, \big)\,,
\\[3mm]
\label{eq:fluidCurrent:Def}
\vec{\calj}(\vec{r}) &=& \rho_{0} \int d^{3} \vec{x}\; \vec{V}(\vec{x})\; \delta^{3} \big( \vec{X}(\vec{x}) - \vec{r}\,\big)\, .
\ena
By integrating over the variable $\vec{r}$, one can instantly see that above definitions are direct results of a system with finite number of particles, in the infinite limit. A straightforward but somewhat nontrivial computation leads to the Poisson algebra between the Euler variables $\rho(\vec{r})$ and $\vec{\calj}(\vec{r})$,
\bea
\Big\{ \rho(\vec{r}),\, \rho(\vec{s}) \Big\}  &=& 0\,,
\\[3mm]
\Big\{ \calj_{i}(\vec{r}),\, \rho(\vec{s}) \Big\} 
&=& 
\rho(\vec{r}) \Big[ \partial_{r_{i}} \delta^{3}( \vec{r} - \vec{s} ) \Big] \,,
\\[3mm]
\Big\{ \calj_{i}(\vec{r}),\, \calj_{j}(\vec{s}) \Big\} 
&=&
\calj_{i}(\vec{s}) \Big[ \partial_{r_{j}} \delta^{3}( \vec{r} - \vec{s} ) \Big] +
\calj_{j}(\vec{r})\Big[ \partial_{r_{i}} \delta^{3}( \vec{r} - \vec{s} ) \Big]  \,.
\ena
These results are completely the same with those obtained in Ref.~\cite{Das:2016hmc,Jackiw:2004nm}. To study the dynamics, we need to have the corresponding Hamiltonian. In Ref.~\cite{Das:2016hmc}, the kinematical part of the Hamiltonian is defined as $H(\vec{r}) = \rho(\vec{r}\,) \vec{v}\,^2(\vec{r}\,)/2$. However, by integrating over the variable $\vec{r}$, one can see that it is not consistent with the infinite limit of the total kinematical energy of a system having finite number of particles. This is also true for the potential energy. Therefore, we will introduce another approach in this section. We introduce following momentum stress tensor,
\bee
\calt_{ij}(\vec{r}) = 
\frac{1}{2}\rho_{0} \int d^{3} \vec{x}\; V_{i}(\vec{x}) V_{j}(\vec{x}) \,
\delta^{3} \big( \vec{X}(\vec{x}) - \vec{r}\, \big) \,.
\ene
We further define the kinematic energy density as the trace of $T_{ij}(\vec{r})$, \ie,
\bee
\calt(\vec{r}) = 
\frac{1}{2}\rho_{0} \int d^{3} \vec{x}\; \vec{V}(\vec{x}) \cdot \vec{V}(\vec{x}) \; 
\delta^{3} \big( \vec{X}(\vec{x}) - \vec{r}\, \big) \,.
\ene
By integrating over the variable $\vec{r}$ one can see that it is consistent with the total kinematical energy of a system having finite number of particles,
\bee
H_{K}=\int d^{3}\vec{r}\;\calt(\vec{r}) 
=
\frac{1}{2} \rho_{0} \int d^{3}\vec{x}\; \vec{V}( \vec{x}) \cdot \vec{V}( \vec{x}) 
\longrightarrow \sum_{i} \frac{1}{2} m\, \vec{V}_{k} \cdot \vec{V}_{k}
\,.
\ene
For the potential energy density we define
\bee
\calu(\vec{r}) = \rho_{0}
\int d^{3} \vec{x}\; U\big( \vec{X}(\vec{x}) \big) \; 
\delta^{3} \big( \vec{X}(\vec{x}) - \vec{r}\, \big)
= \rho(\vec{r})\,U(\vec{r}) \,.
\ene
By integrating over the variable $\vec{r}$ one obtain
\bee
H_{P}=
\int d^{3}\vec{r}\;\calu(\vec{r}) 
=
\rho_{0} \int d^{3}\vec{x}\; U\big( \vec{X}(\vec{x}) \big) \longrightarrow \sum_{i} m \, U\big( \vec{X}_{i} \big)  \,,
\ene
which is again consistent with the total kinematical energy of a system having finite number of particles (the constant factor $m$ is irrelevant, it appears since we use different normalization from the one in Eq.~\eqref{eq:inftyHamiltonian}). Therefore, here and after we will use following Hamiltonian for deriving the fluid dynamics in the Euclidian description,
\bee
H = \int d^{3}\vec{r}\; \Big( \calt(\vec{r}) + \calu(\vec{r}) \Big)\,.
\ene
As a consequence, the equation of motion of the mass density is given as
\bee
\dot{\rho}(\vec{s}) = - \vec{\partial}_{\vec{s}} \cdot \vec{\calj}(\vec{s}) \,.
\ene
Furthermore, the equation of motion of the current density is 
\bee
\dot{\calj}_{i}(\vec{s}) =  - \partial_{s_{j}} \calt_{ij}(\vec{s}) - \rho(\vec{s})\, \partial_{s_{i}} U(\vec{s}) \,.
\ene
This is the momentum equation of motion in conservative form, and clearly it has completely the same expression form with the ones in Lagrangian description. The first term in the right-hand side represents the change in the $i$th component of momentum due to a mismatch in `$i$-momentum' carried through a cell in each of the three orthogonal directions, and the second term stands for the external force which is $\rho(\vec{s})\,\vec{g}(\vec{s})$ for gravity field.

\section{Noncommutative Fluid Variables and Brackets}\label{sec:ncFluid}
In last section we have introduced a new approach for deriving the dynamical equations of the fluid by using the Poisson bracket. In this section, based on this formalism, we extended the fluid dynamics onto noncommutative space which is represented by the algebra in Eq.~\eqref{eq:ncdefine} in noncommutative quantum mechanics. The noncommutative algebra among the canonical Lagrangian variables can be encoded by following extended brackets,
\bea
\Big\{ \underline{V}_{i}( \vec{x}),\, \underline{X}_{j}( \vec{y}  ) \Big\} &=& \frac{1}{\rho_{0}} \delta_{ij} \delta^{3}( \vec{x} - \vec{y} )\,,
\\[3mm]
\Big\{ \underline{X}_{i}( \vec{x}),\, \underline{X}_{j}( \vec{y}  ) \Big\} &=& \frac{1}{\rho_{0}} \theta_{ij} \delta^{3}( \vec{x} - \vec{y} )\,,
\\[3mm]
\Big\{ \underline{V}_{i}( \vec{x}),\, \underline{V}_{j}( \vec{y}  ) \Big\} &=& 0\,,
\ena
where we have used $\vec{\underline{X}}( \vec{x}  )$ and $\vec{\underline{V}}( \vec{x})$ to denote the canonical Lagrangian variables in a noncommutative space. 

Because the variables $\vec{\underline{X}}( \vec{x}  )$ and $\vec{\underline{V}}( \vec{x})$ are not directly measurable, to define an observable, for instance the mass density $\rho(\vec{r})$, in terms of the noncommutative canonical variables, we need a representation of these noncommutative variables in terms of the ordinary ones. Such a representation is given as follows
\bea
\underline{X}_{i}( \vec{x}) &=& X_{i}( \vec{x}) + \frac{1}{2} \theta_{ij}V_{j}( \vec{x})\,,
\\[3mm]
\underline{V}_{i}( \vec{x}) &=& V_{i}( \vec{x}) \,.
\ena
For the Dirac function that frequently appears in the map introduced above, we define its representation as follows
\bee
\delta^{3} \big( \underline{X}_{i}(\vec{x}) - r_{i}\, \big) 
= \delta^{3} \big( X_{i}(\vec{x}) - r_{i}\, \big)
- \frac{1}{2}\theta_{ij}V_{j}(\vec{x})\Big[ \partial_{r_{i}}\delta^{3} \big( X_{i}(\vec{x}) - r_{i}\, \big) \Big]\,.
\ene
The above definition ensure that the noncommutative correction on the Dirac function is consistent in physics after integrating over the space. For instance the mass density
\bee
\int d^{3} \vec{r} \, \underline{\rho}(\vec{r}) 
= 
\int d^{3} \vec{r} \, \rho(\vec{r}) 
- \frac{1}{2} \int d^{3} \vec{r} \; \partial_{r_{i}} \,\theta_{ij} \, J_{j}(\vec{r})
= 
\int d^{3} \vec{r} \, \rho(\vec{r}) \,.
\ene
Here in the second step, the correction disappear because it is a total derivative. This means the number of particles does not change on the noncommutative space, which is consistent the physical requirements. Furthermore, the evolution equation for observable $\calo(\vec{s})$ is defined as follows,
\bee
\dot{\calo}(\vec{s}) 
=
\Big\{ H(X_{i} + \theta_{ij}V_{j}, V_{j} ),\, \calo( \vec{s}; X_{i}, V_{j} ) \Big\}\,.
\ene

However, there is another approach which can avoid to define the representations of the noncommutative quantities. In this approach, the noncommutative effects are taken into account by using the Groenwald-Moyal $\star$-product~\cite{Seiberg:1999vs,Douglas:2001ba,Gangopadhyay:2014afa,Banerjee:2009gr,Holender:2011px,Abdalla:2012tt}, see Eq.~\eqref{eq:starproduct}. The corresponding noncommutative Poisson algebra is defined as 
\bea
\Big\{ V_{i}( \vec{x}),\, X_{j}( \vec{y}  ) \Big\}_{\star} &=& \frac{1}{\rho_{0}} \delta_{ij} \delta^{3}( \vec{x} - \vec{y} )\,,
\\[3mm]
\Big\{ X_{i}( \vec{x}),\, X_{j}( \vec{y}  ) \Big\}_{\star} &=& \frac{1}{\rho_{0}} \theta_{ij} \delta^{3}( \vec{x} - \vec{y} )\,,
\\[3mm]
\Big\{ V_{i}( \vec{x}),\, V_{j}( \vec{y}  ) \Big\}_{\star} &=& 0\,.
\ena
The map introduced in last section is clearly preserved in this approach. The noncommutative properties of the Lagrangian variables are transport to the Euclidian variables via the maps defined in Eq.~\eqref{eq:fluidDensity:Def} and Eq.~\eqref{eq:fluidCurrent:Def}. As a consequences, the algebra between the Eulerian variables can be obtained directly,
\bea
\Big\{ \rho(\vec{r}),\, \rho(\vec{s}) \Big\}_{\star} 
&=& 
\theta_{ij} \Big[ \partial_{r_{i}} \rho(\vec{r}) \Big]
\Big[ \partial_{s_{j}} \delta^{3}( \vec{r} - \vec{s} ) \Big] \,,
\\[3mm]
\Big\{ \calj_{i}(\vec{r}),\, \rho(\vec{s}) \Big\}_{\star} 
&=& 
\bigg\{ \rho(\vec{r})\delta_{ik} - \theta_{jk} \Big[ \partial_{r_{j}} \calj_{i}(\vec{r})\Big] \bigg\}
\Big[ \partial_{r_{k}} \delta^{3}( \vec{r} - \vec{s} ) \Big] \,,
\\[3mm]
\Big\{ \calj_{i}(\vec{r}),\, \calj_{j}(\vec{s}) \Big\}_{\star}  
&=&
\calj_{i}(\vec{s}) \Big[ \partial_{r_{j}} \delta^{3}( \vec{r} - \vec{s} ) \Big] +
\calj_{j}(\vec{r})\Big[ \partial_{r_{i}} \delta^{3}( \vec{r} - \vec{s} ) \Big]  \,.
\ena
One can see that the noncommutativity of the space does not affect the commutator between current density variables. We have checked that the above two approaches give completely the same results (up to leading order of the nocommutative parameter $\theta$).
The evolution of an observable $\calo(\vec{s})$ is subjected to the following equation,
\bee
\dot{\calo}(\vec{s}) 
= \Big\{ H(X_{i}, V_{j} ),\, \calo( \vec{s};   X_{i}, V_{j}) \Big\}_{\star} \,.
\ene
By using this equation, we can easily obtain the equation of motion of the mass density,
\bee\label{eq:ncFluid:MotDensity}
\dot{\rho}(\vec{s}) = - \vec{\partial}_{\vec{s}} \cdot \vec{\calj}(\vec{s}) 
+ \theta_{ij} \Big[ \partial_{s_{i}} \rho(\vec{s}) \Big]
\Big[ \partial_{s_{j}} U( \vec{s} ) \Big] \,.
\ene
Comparing to the results in Ref.~\cite{Das:2016hmc}, there is an additional term due to the potential. This additional contribution disappear in Ref.~\cite{Das:2016hmc} because $U'(\rho)$ (in their notation) is treated as a $\vec{\underline{X}}$-independent function. For the equation of motion of the current density we have
\bee\label{eq:ncFluid:MotCurrent}
\dot{\calj}_{i}(\vec{s}) 
=  - \partial_{s_{j}} \calt_{ij}(\vec{s}) - \rho(\vec{s}) \partial_{s_{i}} U(\vec{s})
 + \theta_{jk} \Big[ \partial_{s_{j}} \calj_{i}(\vec{s})\Big] \Big[ \partial_{s_{k}} U(\vec{s}) \Big] \,.
\ene
Again there is a contribution that depending the derivatives of the potential.

\section{Cosmological Implications}\label{sec:ncCosm}
Several noncommutative extensions of the standard cosmological models have been proposed~\cite{Marcolli:2009in,Chamseddine:2014nxa,Chamseddine:2007bm,Chamseddine:2006ep}, however, as commented in Ref.~\cite{Das:2016hmc}, these models are based on the modified uncertainty relation, and hence difficult to recognize the full physical effects of the noncommutative geometry in cosmology. While a new approach was introduced in Ref.~\cite{Das:2016hmc}, it is incomplete as we have explained and more serious investigations are in order. In this section, along the line in Ref.~\cite{Das:2016hmc}, based on our formalism we briefly discuss the cosmological implications of the noncommutative fluid dynamics. We will show that we arrive at completely different conclusions from the one in Ref.~\cite{Das:2016hmc}.

We will discuss the noncommutative corrections on the Friedmann equation in the Friedmann? Robertson?Walker (FRW) framework of cosmology. In case of vanishing pressure and zero cosmological constant, the equations of motion of the density and current are
\bea
\dot{\rho} &=& -3H\rho \,,
\\[3mm]
\ddot{a} &=& - \frac{4}{ 3}\pi G \rho a \,,
\ena
where $a$ is the scale factor, $G$ is the Newton's constant, and $H=\dot{a}/a$ is the Hubble parameter. The Friedmann equation obtained by combing these two equations is
\bee\label{eq:Friedmann}
H^{2} = \frac{8}{3} \pi G \rho   - \frac{ k_{0} }{a^{2}} \,,
\ene 
where $k_{0}$ is the constant for initial condition.

Let us study the noncommutative corrections on Eq.~\eqref{eq:Friedmann}. We will use the variable $\vec{z}(t)$ to denote the comoving coordinates, and as usual, the proper coordinate $\vec{r}(t)$ is defined as the product of the comoving coordinates and the scale factor $a(t)$, \ie, $\vec{r}(t) = a(t) \vec{z}(t) $. Then in terms of the comoving velocity, the proper velocity is $\dot{\vec{r}}(t)  = H(t) \vec{r}(t) + a \dot{\vec{z}}(t) $. However, our formalism are written in terms of current density, therefore we need to derive it from its definition, \ie, Eq.~\eqref{eq:fluidCurrent:Def}. A straightforward calculation shows
\bee
\vec{\calj}(\vec{r}) = H(t) \, \vec{r}(t) \, \rho(\vec{r}) + a(t)\,\cali(\vec{r})\,,
\ene
where $\cali(\vec{r})$ is the comoving current density. As usual, we assume that the comoving frame is stable, \ie, $\dot{\vec{x}}(t)=0$, and hence the comoving current $\cali(\vec{r})$ vanishes. Therefore the proper current density is reduced to following form,
\bee
\vec{\calj}(\vec{r}) = H(t) \, \vec{r}(t) \, \rho(\vec{r}) \,.
\ene
Similarly one can obtain the momentum stress tensor in this case,
\bee
\calt_{ij}(\vec{r}) = H^{2}(t) \, r_{i}(t) \, r_{j}(t) \, \rho(\vec{r}) \,.
\ene
Inserting these results into the right-hand side of the noncommutative corrected equations of motion of density and current, Eq.~\eqref{eq:ncFluid:MotDensity} and Eq.~\eqref{eq:ncFluid:MotCurrent}, respectively, one has,
\bea
\label{eq:ncFluid:MotDensity:S}
\dot{\rho}(\vec{s}) 
&=& - H \Big[ 3 + \vec{s}\cdot \vec{\partial}_{\vec{s}}  \Big] \rho(\vec{s})
+ \theta_{ij} \Big[ \partial_{s_{i}} \rho(\vec{s}) \Big]
\Big[ \partial_{s_{j}} U( \vec{s} ) \Big] \,,
\\[3mm]
\label{eq:ncFluid:MotCurrent:S}
\dot{\calj}_{i}(\vec{s}) 
&=& - M_{i}(\vec{s}) 
- \rho(\vec{s}) \partial_{s_{i}} U(\vec{s})
 + \theta_{jk} H N_{ij}(\vec{s})\Big[ \partial_{s_{k}} U(\vec{s}) \Big] \,,
\ena
where
\bea
M_{i}(\vec{s}) &=& H^{2} s_{i} \Big[  4  +   \vec{s} \cdot \, \vec{\partial}_{\vec{s}}   \Big] \,\rho(\vec{s})\,,
\\[3mm]
N_{ij}(\vec{s}) &=&  \Big[ \delta_{ij} + s_{i} \partial_{s_{j}} \Big] \rho(\vec{s}) \,.
\ena
In the above derivations, we have assumed that the scale factor $a$ is spatial-independent. However, we will show that in general the noncommutative geometry introduces spatial-dependent corrections on the Friedmann equation. Therefore, all the results below should be understood in perturbative view of point. In spite of this, Eq.~\eqref{eq:ncFluid:MotDensity:S} and Eq,~\eqref{eq:ncFluid:MotCurrent:S} are the master equations describing the noncommutative corrections in case of vanishing comoving current. 

To obtain the explicit noncommutative deformations on the Friedmann equation \eqref{eq:Friedmann}, we need to transform the derivatives with respect to proper coordinates into derivatives with respect to comoving coordinates. This can be easily done by using following relations,
\bea
\partial_{s_{i}} &=& a^{-1}\, \partial_{z_{i}} \,,
\\[3mm]
\partial_{t}\Big|_{\vec{s}} &=& \partial_{t}\Big|_{\vec{z}} - H \Big[\vec{z}\cdot \vec{\partial}_{\vec{z}} \Big] \,.
\ena
After replacing the derivatives in Eq.~\eqref{eq:ncFluid:MotDensity} and Eq.~\eqref{eq:ncFluid:MotCurrent}, we obtain following results,
\bea
\label{eq:cosm:ncEulerRho}
\dot{\rho}(\vec{z}) 
&=& - 3 H \rho(\vec{z}) 
+ a^{-2}\theta_{ij} \Big[ \partial_{z_{i}} \rho(\vec{z}) \Big]
\Big[ \partial_{z_{j}} U( \vec{z} ) \Big] \,,
\\[3mm]
\label{eq:cosm:ncEulerCur}
\dot{\calj}_{i}(\vec{z}) - B_{i}(\vec{z})
&=& - M_{i}(\vec{z})  - a^{-1}\rho(\vec{z}) \partial_{z_{i}} U(\vec{z})
 + a^{-2}\dot{a}\, \theta_{jk} N_{ij}(\vec{z}) \Big[ \partial_{z_{k}} U(\vec{z}) \Big] \,,
\ena
where
\bea
\label{eq:ncCurrent:Time}
\dot{\calj}_{i}(\vec{z}) &=& \Big[  \ddot{a} \rho(\vec{z}) + \dot{a}  \, \dot{\rho}(\vec{z}) \Big] \, z_{i}\,,
\\[3mm]
B_{i}(\vec{z}) &=& H \dot{a} z_{i} \Big[ 1 + \vec{z}\cdot \vec{\partial}_{\vec{z}}  \Big] \,\rho(\vec{z})\,,
\\[3mm]
M_{i}(\vec{z}) 
&=&
H \dot{a}  z_{i} \,\Big[ 4 +  \vec{z}\cdot \partial_{\vec{z}}   \Big] \,\rho(\vec{z}) \,,
\\[3mm]
N_{ij}(\vec{z}) &=&  \Big[ \delta_{ij} + z_{i} \partial_{z_{j}} \Big] \rho(\vec{z}) \,.
\ena
Applying the equation of motion of the mass density, \ie, Eq.~\eqref{eq:cosm:ncEulerRho}, on Eq.~\eqref{eq:ncCurrent:Time}, the equation of motion of the current density, \ie, Eq~\eqref{eq:cosm:ncEulerCur}, can be further simplified as follows,
\bee\label{eq:cosm:ncEulerCurrSim}
\ddot{a}\,  z_{i}
=   - a^{-1} \partial_{z_{i}} U(\vec{z}) + \dot{a}  \, \theta_{ij} \big[ \partial_{z_{j}} U( \vec{z} ) \big] \,.
\ene
We can clearly see that the spatial variations of the potential play crucial role in the noncommutative corrections. Here we discuss the physical consequences of a dust potential restricted into following form,
\bee\label{eq:potential}
U( \vec{z} ) = \frac{2}{3} \pi G \rho(\vec{z})  (a \vec{z})^{2} \,.
\ene
Inserting this potential into the modified equations of motion, Eq.~\eqref{eq:cosm:ncEulerRho} and Eq.~\eqref{eq:cosm:ncEulerCurrSim}, respectively, we have
\bea
\label{eq:cosm:ncEulerRho:dust}
\dot{\rho}(\vec{z}) 
&=& - 3 H \rho(\vec{z}) 
- \frac{4}{3} \pi G \rho(\vec{z}) \, K_{\theta}(\vec{z}) \,,
\\[3mm]
\label{eq:cosm:ncEulerCur:dust}
 \ddot{a}   
&=& - \frac{4}{3} \pi a G \rho(\vec{z}) 
\left[   1 +  \frac{ K(\vec{z}) - a\,\dot{a}\, K_{\theta}(\vec{z}) }{ 2\rho(\vec{z}) }  \right]  \,, 
\ena
where
\bea
\label{eq:cosm:Kfunc}
K(\vec{z}) &=&  \vec{z} \cdot \vec{ \partial}_{ \vec{z}} \,\rho(\vec{z}) \,,
\\[3mm]
\label{eq:cosm:Kncfunc}
K_{\theta}(\vec{z}) &=&  \vec{\theta} \cdot \Big[ \vec{z} \times \vec{ \partial}_{ \vec{z}} \,\rho(\vec{z})  \Big] \,,
\ena
where we have used the notation $\theta_{ij} = \epsilon_{ijk}\theta_{k}$ with $\epsilon_{ijk}$ the totally anti-symmetric tensor. It is clear that if the inhomogeneity and anisotropy of the density $\rho(\vec{z})$ are small, the two functions $K(\vec{z})$ and $K_{\theta}(\vec{z})$ are variations of $\rho(\vec{z})$ under translation and rotation around the direction $\vec{\theta}$, respectively. This property is important for observing the noncommutative corrections. For instance, if the potential $U(\vec{z})$ is spherically symmetric, which implies that the mass density $\rho(\vec{z})$ is also spherically symmetric, then $K_{\theta}(\vec{z})=0$ due to that its partial derivatives with respect to polar and azimuthal angles vanishes. In order to see the full results, let us assume that $U(\vec{z})$ is not spherically symmetric for the moment. Then by combing Eq.~\eqref{eq:cosm:ncEulerRho:dust} and Eq.~\eqref{eq:cosm:ncEulerCur:dust} one has,
\bee
\frac{1}{2} \frac{\partial}{\partial t} (\dot{a}^2) 
= 
\frac{4}{3} \pi G \left[   1 +  \frac{ K(\vec{z}) - a\,\dot{a}\, K_{\theta}(\vec{z}) }{ 2\rho(\vec{z}) }  \right]  
 \left[ \partial_{t} \left[ a^2 \rho(\vec{z})  \right] + \frac{4}{3} \pi G \rho(\vec{z}) \, K_{\theta}(\vec{z}) \right] \,.
\ene
Then the original Friedmann equation receives noncommutative corrections,
\bee
H^{2} = \frac{8}{3} \pi G \rho - \frac{ k_{NC} }{a^{2}} \,,
\ene
where $k_{NC} = k + k_{\theta}$, and
\bea
\label{eq:k}
k &=& k_{0} - 
\frac{4}{3} \pi G\int dt \, \frac{ K(\vec{z})  }{ \rho(\vec{z}) } \frac{\partial}{\partial t} \left[ a^2 \rho(\vec{z})  \right] \,,
\\[3mm]
k_{\theta} &=& \frac{4}{3} \pi G\int dt \left\{ 
\left[ a\dot{a} \frac{ K_{\theta}(\vec{z}) }{ \rho(\vec{z}) } \right] 
\frac{\partial}{\partial t} \left[ a^2 \rho(\vec{z})  \right] 
-   \frac{4}{3} \pi G \rho(\vec{z}) \, K_{\theta}(\vec{z})\left[   1 +  \frac{ K(\vec{z}) }{ 2\rho(\vec{z}) } \right]
\right\}\,.
\ena
Note that, comparing to the results in Ref.~\cite{Das:2016hmc}, there is an additional term in Eq.~\eqref{eq:k} which is $\theta$-independent. This part is denoted by $\phi_{pec}$ in the potential, and was dropped. But here, because we can not judge if those contribution is smaller then the noncommutative corrections, therefore we keep all the terms. One can see clearly that the noncommutative space can generally introduce {\em additional} anisotropy and inhomogeneity represented by the functions $K(\vec{z})$ and $K_{\theta}(\vec{z})$. These two functions measure non-trivial distributions of the density function $\rho(\vec{z})$ in different directions: $K(\vec{z})$ represents the inhomogeneity of the density distribution, $K_{\theta}(\vec{z})$ characterizes the anisotropy under rotation around the direction $\vec{\theta}$. However, one should note that the non-trivial corrections depend essentially on the original density distribution function $\rho(\vec{z})$, as can be seen in Eq.~\eqref{eq:cosm:Kfunc} and Eq.~\eqref{eq:cosm:Kncfunc}. And for a spherically symmetric mass density there is no noncommutative correction. Nevertheless, in general case, because $K(\vec{z})$ and $K_{\theta}(\vec{z})$ measure the inherent inhomogeneity and anisotropy of the density distribution in different directions, respectively, the noncommutative corrections become important and can be clearly distinguished from the original one.

\section{Summary and Conclusions}\label{sec:sum}
In summary we proposed a refined approach for studying the fluid dynamics on noncommutative space. We start from the single particle picture, and then the map from canonical Lagrangian variables to Eulerian variables are obtained in the infinite limit of a system with finite number of particles. This approach makes sure that both the kinematical and potential energies are taken into account correctly, and the equations of motion of the mass density and current density are naturally expressed into conservative form.

Based on this, we extended the usual Poisson bracket (for single particle) to the case with noncommutative algebra, and in the infinite limit the algebra among Eulerian variables are obtained. Furthermore, we find that the noncommutative corrections on the equations of motion are generally depend on the derivatives of potential.

Most interestingly, applying our results for the Friedmann equation, we find that the noncommutative algebra does modify the usual Friedmann equation. This conclusion is completely different from the one in Ref.~\cite{Das:2016hmc}. This is because we treat the potential in different ways, as we have explained in the text. We briefly studied the influences of these equations on the ordinary Friedmann equation with potential given in Eq.~\eqref{eq:potential}. We find that noncommutative space can generally introduce {\em additional} anisotropy and inhomogeneity represented by the functions $K(\vec{z})$ and $K_{\theta}(\vec{z})$. These two functions measure the symmetry properties of the density function $\rho(\vec{z})$ under translation and rotation around the direction $\vec{\theta}$, respectively, and hence $K(\vec{z})$ refers to the inhomogeneity and $K_{\theta}(\vec{z})$ characterizes the anisotropy. However, as expected, the noncommutative corrections  depend essentially on the derivatives of the density distribution function $\rho(\vec{z})$, as can be seen in Eq.~\eqref{eq:cosm:Kfunc} and Eq.~\eqref{eq:cosm:Kncfunc}. Furthermore, for spherically symmetric mass density and potential the noncommutative corrections vanish. Nevertheless, in general case, because $K(\vec{z})$ and $K_{\theta}(\vec{z})$ measure the inherent inhomogeneity and anisotropy of the density distribution in different directions, respectively, the noncommutative corrections become important and can be  distinguished from the usual distribution.

\section*{Acknowledgements}
K. M. is supported by the China Scholarship Council, and the National Natural Science Foundation of China under Grant No. 11647018 and 11705113, and partially by the Natural Science Basic Research Plan in Shaanxi Province of China under Grant No. 2017JM1032.

\bibliography{aString}

\end{document}